\documentclass[12pt]{article}

\usepackage{epsfig}
\usepackage{amsmath}
\usepackage{amsfonts}
\usepackage{mathrsfs}
\usepackage{units}

\newcommand{\beq}{\begin{equation}}
\newcommand{\eeq}[1]{\label{#1}\end{equation}}
\newcommand{\eeqn}{\end{equation}}

\newcommand{\leqn}[1]{\eqref{#1}}

\def\to{\rightarrow}
\def\tr{{\rm tr}}

\newcommand{\lagrange}[1]{\mathscr{#1}}   
\newcommand{\sptwo}{B{}}           
\newcommand{\spother}[1]{\gamma^*_{#1}{}}   
\newcommand{\fst}{H}                      
\newcommand{\abs}[1]{\lvert #1 \rvert}    
\newcommand{\rselwr}[3]{#1^{#2}_{\phantom{#2}#3}}  
\newcommand{\lwrrse}[3]{#1_{#2}^{\phantom{#2}#3}}  
\newcommand{\set}[1]{\mathbb{#1}}         
\newcommand{\warp}[1]{\ensuremath{e^{#1 k \abs{\smash[b]{y}}}}}  
\newcommand{\ud}{\mathrm{d}}              
\newcommand{\kkn}[1]{#1^{(n)}}            
\newcommand{\kkm}[1]{#1^{(m)}}            

\def\LIR{\Lambda_{\rm IR}}

\DeclareMathOperator{\sgn}{sgn}

\def\stacksymbols #1#2#3#4{\def\theguybelow{#2}
    \def\vp{\lower#3pt}
    \def\sp{\baselineskip0pt\lineskip#4pt}
    \mathrel{\mathpalette\intermediary#1}}

\def\intermediary#1#2{\vp\vbox{\sp
     \everycr={}\tabskip0pt
     \halign{$\mathsurround0pt#1\hfil##\hfil$\crcr#2\crcr
              \theguybelow\crcr}}}

\def\gsim{\stacksymbols{>}{\sim}{2.5}{.2}}
\def\lsim{\stacksymbols{<}{\sim}{2.5}{.2}}
\begin{document}

\begin{titlepage}

\vskip.5cm
\begin{center}
{\huge \bf Tensor Reggeons from Warped Space at the LHC} \\
\vskip0.4cm
{\huge \bf } 
\vskip.2cm
\end{center}
\vskip1cm

\begin{center}
{\bf Maxim Perelstein and Andrew Spray} \\
\end{center}
\vskip 8pt

\begin{center}
	{\it Institute for High Energy Phenomenology\\
	Newman Laboratory of Elementary Particle Physics\\
	Cornell University, Ithaca, NY 14853, USA } \\

\vspace*{0.1cm}

{\tt  mp325@cornell.edu,
aps37@cornell.edu}
\end{center}

\vglue 0.3truecm

\begin{abstract}
\vskip 3pt \noindent
The hierarchy problem can be addressed by extending the four-dimen\-sional space-time to include an extra compact spatial dimension with non-trivial ``warped" metric, as first suggested by Randall and Sundrum. If the Randall-Sundrum framework is realized in string theory, the effective value of the string scale in the vicinity of the infrared boundary should be in the TeV domain. The most attractive models of this type embed the Standard Model particles as zero-modes of five-dimensional fields. In such models, Regge excitations of the Standard Model states should appear around the TeV scale. We construct a toy model that describes tensor (spin-2) excitations of the Standard Model gauge bosons, and their on-shell couplings with light matter and gauge fields, within this framework. We use this toy model to predict the phenomenologically important features of the tensor Regge gluon, such as its mass, production cross section at the LHC, and decay patterns.  
\end{abstract}

\end{titlepage}

\section{Introduction}

In the last decade, a number of interesting new physics scenarios involving extra dimensions of space relevant at the TeV scale have been proposed. Two of the best-known examples are the models with flat Large Extra Dimensions (LED), due to Arkani-Hamed, Dimopoulos, and Dvali~\cite{ADD}, and the models with a single extra dimension with a non-factorizable (``warped'') metric, suggested by Randall and Sundrum (RS)~\cite{RS}. Both classes of models address the gauge hierarchy problem, motivating them as complements for the SM at the weak scale (although in the LED model, an additional mechanism is needed to stabilize the radii of the extra dimensions at large values in natural units). In  the case of LED, this is achieved by bringing the fundamental scale of quantum gravity down into the TeV domain. If the LED scenario is realized, and if string theory serves as the ultraviolet completion of Einstein's general relativity, the stringy nature of the SM particles should become apparent at the TeV scale. In particular, the upcoming experiments at the Large Hadron Collider (LHC) could observe inherently stringy Regge excitations of the familiar SM states.  A phenomenological study of collider signatures of these states, based on simple toy models embedding parts of the SM into string theory, was initiated by one of us (MP), in collaboration with Cullen and Peskin, in Ref.~\cite{CPP}. A large body of literature exists on this subject; see, e.g., Refs.~\cite{Hunter}, \cite{Regge_others} for some examples, and Ref.~\cite{Lust_review} for a recent review. 

The RS model, viewed as a five-dimensional theory, resolves the hierarchy problem in a similar way: while the fundamental 5D Planck scale $M^*_{\rm Pl}$ is close to its 4D value (of the order of $10^{19}$ GeV), the actual scale where gravitational physics becomes strongly coupled depends on the position in the extra dimension, due to the non-trivial dependence of the metric on this coordinate. In particular, the scale near the ``infrared (IR) brane'', where the Higgs field is localized, is in the TeV domain, and that is where the Higgs loop divergences are cut off. If the RS setup emerges as part of the compactification manifold in a weakly-coupled string theory, the fundamental 5D\footnote{We will not be concerned with the compactification of the other 5 dimensions at this point, assuming for simplicity that their radii are of order inverse $M^*_{\rm Pl}$ and thus they can be safely integrated out.} string scale, $M_S^*$, should lie parametrically below $M^*_{\rm Pl}$, and parametrically above the curvature scale $k$:
\beq
k \ll M_S^* \ll M^*_{\rm Pl}.
\eeq{scales} 
Phenomenologically, no large hierarchy can exist between $k$ and $M^*_{\rm Pl}$ (increasing their ratio would exacerbate the already non-trivial tension between experimental constraints on the model and the fine-tuning in the Higgs mass); values of order 10 are preferred. The inequality~\leqn{scales} then implies $M_S^*\sim 10^{18}-10^{19}$ GeV.\footnote{Another obvious implication is that no large separation of scales is possible, and the approximation of weakly-coupled strings propagating on a smooth geometric background is probably subject to sizable corrections. Since in this paper our goal is to build a toy model to describe the major features of Reggeon phenomenology, rather than a rigorous calculation, we will not be concerned about this point.} The Regge excitations of the states that are free to propagate in the full 5D space will appear as 5D fields, with mass terms in the 5D lagrangian of order $M_S^*$. However, in the neighborhood of the IR brane, the masses will be warped down to the TeV scale, and upon Kaluza-Klein (KK) decomposition we should expect to see 4D Reggeons with masses in the TeV domain. Although the original RS model had all of the SM fields confined on the IR brane, it was subsequently realized that a model with the full SM (with the possible exception of the Higgs) propagating in the full 5D space is more interesting: it can naturally explain the apparent unification of gauge couplings~\cite{RSgut}, avoid precision electroweak constraints~\cite{ADMS}, and has attractive mechanisms to explain the fermion mass hierarchy~\cite{GN} and suppression of flavor-changing neutral currents~\cite{APS}. This setup has also been used to construct ``Higgsless" models~\cite{HL}, where electroweak symmetry is broken by boundary conditions on the 5D gauge fields.  
In these models, one expects a rich Reggeon sector to appear around the warped-down string scale. The goal of this paper is to construct a simple toy model incorporating some of the main features of this sector relevant for collider searches, and to discuss the resulting phenomenology.

Before proceeding, let us comment on how the Regge physics appears in the dual four-dimensional picture. In this picture, the warped-down Planck scale is the scale at which conformal invariance of the fundamental gauge theory is broken. The Higgs and all other states localized at, or near, the IR brane of the 5D model, can be understood as bound states of the fundamental gauge degrees of freedom, with binding energies of order TeV. The Regge states described by our toy model are no exception: from the dual point of view, they are simply higher-spin bound states (e.g., the first Regge excitations of gauge bosons are spin-2 ``glueballs''). In principle, both descriptions can provide interesting information. However, for the low-lying Regge states (below the warped-down Planck scale) that are our focus here, the five-dimensional description is clearly advantageous, since in it these states are weakly coupled. 

The studies of Regge phenomenology in LED models are based on a well-known result in string theory: the Veneziano amplitudes for tree-level scattering of open-string states. Factorizing these amplitudes on their poles determines the Reggeon masses and their (on-shell) couplings to the SM states, which is sufficient to model their collider signatures. Unfortunately, the Veneziano amplitudes only apply to strings propagating on backgrounds with flat (Minkowski) metric, and their generalization to warped spaces such as the RS model is presently unknown. Therefore, we will pursue a different approach. We will restrict our attention to a small subset of the Regge states, namely, the lowest-lying spin-2 Regge excitations of the SM gauge bosons, in particular SM gluon. These states present most realistic targets for collider searches, due to a possibility of relatively large production cross sections at hadron colliders, and their higher-spin nature would make them striking signatures for low-scale string theory. We will begin, in Section 2, by constructing a Lagrangian which reproduces their masses and on-shell couplings to SM in flat space, as obtained from Veneziano amplitudes in previous work. We will then generalize this Lagrangian, in Sections 3 and 4, to spaces with arbitrary metric, using the standard trick of introducing metric factors and covariant derivatives to restore general covariance. (In fact, a slightly non-minimal extension will be preferred, in order to maintain a simple form of the gauge invariance for spin-2 states.) In Section 3, we will also study the KK decomposition of a massive 5D spin-2 field, which to the best of our knowledge has not yet been considered in the literature.
In Section 5, we will outline the predictions of our model for the LHC  phenomenology of the 4D tensor Regge excitation of the gluon.\footnote{High-energy behavior of scattering amplitudes in RS space has been considered in Ref.~\cite{Meade}. An approach similar to ours has been applied recently in Ref.~\cite{HMRR} to spin-3/2 Regge excitations of the top quark.}

\section{A Model for Reggeons in Flat 4D Space}
\label{sec:4D}

Our starting point is the toy model proposed by Cullen, Perelstein and Peskin (CPP) in Ref.~\cite{CPP}. This toy model embeds QED of electrons and photons and QCD of quarks and gluons into string theory as zero modes of open strings living on coincident D3 branes.  Factorizing the tree-level scattering amplitudes between these states at their $s$-channel poles at $s=M_S^2$ provides their (on-shell) couplings to the first-level Reggeons. Our goal in this section is to encode these couplings in a Lagrangian, which can then be generalized to the Randall-Sundrum model. 

\subsection{Stringy Toy Model of Electrons and Photons} 

To describe a string embedding of electrons and photons, the CPP model introduces two coincident D3 branes.  The low-energy physics of this configuration is given by an ${\cal N}=4$ supersymmetric theory with a $U(2)$ gauge group. However, if the external states are chosen from a restricted set consisting of a single (diagonal) gauge boson and two (off-diagonal) gauginos, the internal propagators in any tree-level diagram must also come from this set. The gauge boson is identified with the photon and the two gauginos with the two helicity states of the electron. Taking the low-energy limit of the tree-level scattering amplitudes of this string theory reproduces the familiar helicity amplitudes of QED, while at high energies the amplitudes exhibit the Regge poles characteristic of string theory. In particular, on-shell couplings of the string Regge resonances to the SM (zero-mode) fields can be obtained by factorizing the amplitudes on the $s$-channel Regge poles. 

Since the kinematic reach of near-future collider experiments is unlikely to extend deep into the Regge domain, we will focus our analysis on modelling the phenomenology of the first Regge level. Moreover, as the first step, we will restrict ourselves to the excitations of the SM gauge bosons. These states can be singly produced in the collision of SM fermion-antifermion pairs, as well as, in the case of the $SU(3)$ Reggeons, SM gluons. In the CPP model, the bosonic states at the first Regge level are a spin-2 state $\gamma^*_2$, a spin-1 state $\spother{1}$, and four spin-0 states $\spother{0}^{(i)}$. We will focus on the spin-2 states in this paper, since they would provide the most unambiguous signature of stringy physics if discovered; the approach of this paper, however, can be easily generalized to include the lower-spin states. Our first task is to construct a field theory model to reproduce the Feynman rules for the couplings of these states to SM, derived in Ref.~\cite{CPP}. We introduce Reggeon field $B_{\mu\nu}(x)$. (Since open strings are confined to D3 branes, it is a 4D field.) The quadratic action has the usual form:
\beq
{\cal L}_{S=2} \,=\, \frac{1}{4} \fst^{\lambda\mu\nu} \fst_{\lambda\mu\nu} - \frac{1}{2} \rselwr{\fst}{\lambda\mu}{\mu} \lwrrse{\fst}{\lambda\nu}{\nu} + \frac{1}{2} m^2 \left\{ \left( \lwrrse{\sptwo}{\mu}{\mu} \right)^2 - \sptwo^{\mu\nu} \sptwo_{\mu\nu} \right\} \,,
\eeq{Bin4D}
where we introduced the field-strength tensor  $\fst_{\lambda\mu\nu} \equiv \partial_\lambda \sptwo_{\mu\nu} - \partial_\mu \sptwo_{\lambda\nu}$, and $m\equiv M_S$ is the Reggeon mass.  The kinetic term is, up to a factor, the same as the standard graviton action found by expanding the Einstien-Hilbert action to quadratic order in $h_{\mu\nu} \equiv g_{\mu\nu} - \eta_{\mu\nu}$.  The mass term has the Fierz-Pauli form~\cite{fierzpauli} that is necessary for unitarity.

The interactions of SM electrons and photons (string zero-modes) with the Regge states can be described by the following lagrangian: \begin{multline}
\lagrange{L} _{\rm int}= \frac{i \, e}{\sqrt{2}M_S}\,\left( \partial^\mu \bar{\psi} \gamma^\nu \psi - \bar{\psi} \gamma^\nu \partial^\mu \psi \right) \sptwo_{\mu\nu} \\ \,+\, \frac{e}{\sqrt{2}M_S}\,\left( F^{\rho\mu} \rselwr{F}{\nu}{\rho} - \frac{1}{4} F^{\rho\sigma}F_{\rho\sigma}\eta^{\mu\nu}\right) \sptwo_{\mu\nu} 
+  {\rm~(vectors, scalars)}
\label{Leff4} \end{multline}
where $\psi$ is the electron field and $F$ is the electromagnetic field strength. This lagrangian can be read off from the Feynman rules in Fig.~7 of Ref.~\cite{CPP}. Note that the Feynman rules were derived by factorizing the Veneziano amplitudes on the Regge poles, and thus only contain information about interactions of on-shell particles. So, the model~\leqn{Leff4} is only valid on-shell: there may be additional operators not included here that vanish for on-shell particles. It is adequate for describing resonant production of Regge states in SM collisions at tree level, which should be sufficient for understanding the main features of their collider phenomenology.  

\subsection{Stringy Toy Model of Quarks and Gluons} 
\label{sec:stringyQCD}

The Regge gluon is of great interest phenomenologically, since it is strongly interacting and could have a large production cross section at hadron colliders. The quark-antiquark-Regge gluon coupling is simply obtained from the $e^+e^-\gamma^*$ vertex by replacing $e\rightarrow g$, promoting derivatives $\partial_\mu$ to covariant derivatives $D_\mu$, and introducing the usual color structure~\cite{CPP}: 
\beq
{\cal L}_{q\bar{q}g^*} = \frac{i g}{\sqrt{2}M_S} \,\left( (D^\mu \bar{q}) \gamma^\nu  \tilde{\sptwo}_{\mu\nu} q - \bar{q} \gamma^\nu \tilde{\sptwo}_{\mu\nu} D^\mu q \right) + {\rm (vectors, scalars)}\,.
\eeq{Lqcd4}  
Here we defined $\tilde{\sptwo}_{\mu\nu}=\sptwo_{\mu\nu}^a t^a$, where $\sptwo_{\mu\nu}^a~(a=1\ldots 8)$ is the Regge gluon field, and $t^a$ are the fundamental representation generators of QCD $SU(3)$, normalized by ${\rm tr}(t^a t^b)=\delta^{ab}/2$. Note that the Regge gluon field transforms linearly in the adjoint representation of $SU(3)$: 
\beq
\tilde{\sptwo}_{\mu\nu} \rightarrow {\cal U}\,\tilde{\sptwo}_{\mu\nu} 
\,{\cal U}^{-1},~~~{\cal U} = \exp\left(it^a\theta^a\right)\,.
\eeq{Adj}
Since $q\rightarrow {\cal U}q$, this ensures the gauge invariance of the coupling~\leqn{Lqcd4}. 

In the Randall-Sundrum model, the Regge gluon wavefunctions are localized near the TeV brane (as will be shown below), while the wavefunctions of light fermions may be localized at the opposite ``Planck'' boundary. In this case, the coupling of the Regge gluon to light SM quarks is strongly suppressed, and the most important production channel for $g^*$ is via gluon fusion. (The zero-mode gluon wavefunction is constant across the extra dimension.) To model this interaction, we need to obtain the gluon-gluon-Regge gluon vertex in the CPP model. Since this was not done in Ref.~\cite{CPP}, let us briefly outline the derivation here. The CPP model identifies gluons with open strings ending on a stack of 4 coincident D3 branes. The 4-gluon scattering amplitude is given by
\begin{multline}
{\cal A}(1,2,3,4) \,=\, g^2 A(1,2,3,4) \, {\cal S} (s,t) \, \tr [t^1 t^2 t^3 t^4 + t^4 t^3 t^2 t^1] \\
+  g^2 A(1,3,2,4) \, {\cal S} (u,t) \, \tr [t^1 t^3 t^2 t^4 + t^4 t^2 t^3 t^1] \\+
g^2 A(1,2,4,3) \, {\cal S} (s,u) \, \tr [t^1 t^2 t^4 t^3 + t^3 t^4 t^2 t^1]\,, 
\label{4gluon}\end{multline}
where $t^i\equiv t^{a_i}$ are the generators of the fundamental representation of $SU(3)$ ("Chan-Paton factors"), while
\beq
{\cal S} (s,t) = \frac{\Gamma(1-\alpha^\prime s) \Gamma(1-\alpha^\prime t)}{\Gamma (1- \alpha^\prime s- \alpha^\prime t)}
\eeq{ReggeF}
is the string formfactor (essentially the Veneziano amplitude), and the $A$'s denote the color-ordered four-point gauge theory amplitudes. (Note that $\alpha^\prime = 1/M_S^2$.) At tree level, all non-vanishing color-ordered helicity amplitudes for four-gluon scattering can be obtained from the two basic ones by index permutations. The basic amplitudes are
\beq
A(1^+,2^-,3^-,4^+) = -4\frac{t}{s}\,,\quad A(1^+,2^-,3^+,4^-) = -4\frac{u^2}{st}\,,
\eeq{helamps}
where helicities are directed inward. Using these amplitudes in Eq.~\leqn{4gluon} and factorizing the amplitudes on the Regge pole, $s=M_S^2$, we obtain
\begin{align}
{\cal A} (g^+g^+ \rightarrow g^+g^+) & = - 2 \, g^2 \frac{s}{s-M_S^2}\,\cdot\,{\cal C}^{1234}\,,\\
{\cal A} (g^+g^- \rightarrow g^+g^-) & = - 2 \, g^2 \frac{u^2}{s^2}\,\frac{s}{s-M_S^2}\,\cdot\, {\cal C}^{1234}\,,
\label{amps}\end{align}
All other non-vanishing amplitudes are related to these two by parity. 
Note that the kinematic dependence of the factorized four-gluon amplitudes exactly matches that of the four-photon amplitudes studied 
in Ref.~\cite{CPP}, implying that the Lorentz structure of the $ggg^*$ vertices is the same as for the $\gamma\gamma\gamma^*$ vertices. 
The color factor is given by
\beq
{\cal C}^{1234} = 2 \bigl( \tr [t^1t^2t^3t^4] + \tr [t^4t^3t^2t^1] + \tr [t^1t^2t^4t^3] + \tr [t^3t^4t^2t^1] \bigr) , 
\eeq{Camp} 
where as before $t^i \equiv t^{a_i}$.
To factorize this, we use the well-known $SU(N)$ identity
\beq
2\,\sum_{a=1}^{N^2-1} (t^a)^i_j (t^a)^k_l = \delta^i_l \delta^k_j \,-\, \frac{1}{N} 
\delta^i_j \delta^k_l\,.
\eeq{iden}
We obtain
\begin{multline}
{\cal C}^{1234} = 4\,\sum_{a=1}^{N^2-1} \left( \tr [t^1t^2t^a] + \tr [t^2t^1t^a]\right) 
\,\cdot\,\left( \tr [t^3t^4t^a] + \tr [t^4t^3t^a]\right) 
\\ + \frac{8}{3} \tr[t^1 t^2] \,\cdot\,\tr[t^3t^4]\,. 
\label{CfactorFact}\end{multline}
This suggests that there are in fact 9 Regge gluons propagating in the $s$ channel in four-gluon scattering: a color-octet, coupled with strength $g$, and a color-singlet, coupled with strength $g/\sqrt{3}$. 
The appearance of the color singlet Reggeon in the CPP model was already noted in Ref.~\cite{CPP}; in fact, there is an additional {\it massless} color-singlet vector boson in this model as well, due to an extra $U(1)$ factor in the low-energy theory of strings on D3-branes.
In realistic string models, such $U(1)$ factors are typically anomalous, and the corresponding gauge bosons obtain masses at the string scale via Green-Schwartz mechanism. This mechanism will probably also affect the mass of the color-singlet Reggeon. In general, the fate of this state appears model-dependent, and even if it is present at $M_S$, its effect on phenomenology would be subdominant to the color-octet state due a smaller number of degrees of freedom and a suppressed coupling. Thus, we will focus on the color-octet Reggeon. To obtain the Feynman rules for the interactions of this state with SM gluons, one needs to simply multiply the photon-Regge photon vertices in Fig.~7 of Ref.~\cite{CPP} by a color factor
\beq
C^{abc} \,=\, 2 \left( \tr [t^a t^b t^c] + \tr [t^a t^c t^b] \right)\,,
\eeq{Cfactor}
and substitute $e\rightarrow g$.
The corresponding term in the Lagrangian is
\beq
{\cal L}_{ggg^*} = \frac{g}{\sqrt{2}M_S}\,C^{abc}\,\left( F^{a\rho\mu} F^{b\nu}_\rho - \frac{1}{4} F^{a\rho\sigma}F^b_{\rho\sigma}\eta^{\mu\nu}\right) \sptwo^c_{\mu\nu}\,+ {\rm (vectors, scalars)}\,,
\eeq{Lqcd4gluon}
where $F^a_{\mu\nu} = \partial_\mu A^a_\nu - \partial_\nu A^a_\mu +
i g f^{abc} A^b_\mu A^c_\nu$ is the gluon field strength. As before, it is important to keep in mind that this Lagrangian is only valid for on-shell production of Regge gluons in SM gluon collisions. Using 
\beq
F_{\mu\nu}^a t^a \rightarrow {\cal U}\,(F_{\mu\nu}^a 
t^a)\,{\cal U}^{-1},
\eeq{Fgt}
together with Eq.~\leqn{Adj}, it is easy to show that the coupling~\leqn{Lqcd4gluon} is $SU(3)$ invariant. Note that in order to preserve gauge invariance, the derivatives in the kinetic lagrangian of the Reggeon, Eq.~\leqn{Bin4D}, need to be promoted to covariant derivatives, leading to additional couplings between gluons and the Reggeon. However, these vertices always involve two Reggeon fields, and thus do not contribute to on-shell single-Reggeon production, making them irrelevant for the analysis of this paper.

In the CPP model, the SM quarks were modeled, somewhat naively, as open-string zero modes in adjoint representation of an enlarged gauge group containing the SM $SU(3)$~\cite{CPP}. A more realistic construction was used in Ref.~\cite{Hunter}, where the SM fermions are described by boundary operators in the open string CFT. In practice, as far as the couplings of the first Regge resonance to on-shell quarks are concerned, the two approaches produce identical results. For example, using the couplings~\leqn{Lqcd4} and~\leqn{Lqcd4gluon}, obtained from the CPP model, we reproduce the $s$-channel pole at $s=M_S^2$ of the two-fermion-two-gluon amplitude in Eq.~(5.39) of Ref.~\cite{Hunter}. We do {\it not} reproduce the four-fermion amplitude, since, as remarked in Ref.~\cite{Hunter}, states other that the Reggeon are being exchanged in this channel, leading to highly model-dependent results.  

\section{A Model for Warped-Space Reggeons}

In this section, we generalize the above toy model to Reggeons propagating in the Randall-Sundrum (RS) space, and present the Kaluza-Klein decomposition of a massive spin-2 field in RS.

\subsection{The Randall-Sundrum Orbifold}

Let us first review the relevant features of the Randall-Sundrum (RS) orbifold.  Topologically, the RS space is the direct product of Minkowski space and the $S^1/\set{Z}_2$ orbifold.  We use coordinates $(x^\mu,y)$ where $y$ spans the extra dimension:
\begin{equation}
y \in (-\pi R, \pi R ] ; \qquad y \sim y + 2 \pi R .
\label{RScoords}\end{equation}
The orbifold symmetry takes $y \to -y$; the orbifold fixed points are at $y=0,\pi R$.  The interval in these coordinates is
\begin{equation}
\ud s^2 = \warp{-2} \, \ud x^\mu \, \ud x_\mu - \ud y^2\,.
\label{RSinterval}\end{equation}
The RS space is  
a solution of the Einstein equations with a 5D cosmological constant and 3-branes at the orbifold fixed points. The brane at $y=0$ is referred to as the ultraviolet (UV) brane, while the brane at $y=\pi R$ is the infrared (IR) brane. The curvature scale $k$ is of order (though somewhat below) the 5D Planck scale $M^*_{\rm Pl}$, which in turn is essentially identical to the 4D Planck scale $M_{\rm Pl}$. To solve the hierarchy problem, the model parameters must obey
\beq
e^{\pi k R} \sim \frac{M_{\rm Pl}}{{\rm TeV}} \sim 10^{16}\,.
\eeq{RSpars}
Away from the orbifold fixed points (``in the bulk'') the RS space is isomorphic to AdS$^5$. The non-vanishing Christoffel symbols are
\beq
\Gamma^\mu_{\nu 5}\,=\, -k \sgn (y) \delta^\mu_\nu,~~~\Gamma^5_{\mu\nu} \,=\, -k \sgn(y) e^{-2k|y|} \eta_{\mu\nu}\,.
\eeq{Ch}
In the bulk, the Riemann tensor has the form 
\beq
R^M_{\phantom{M}NKL} = k^2 \left( \delta^M_K g_{NL} - \delta^M_N g_{KL} \right)\,.
\eeq{Riemann}
The discontinuities of the Christoffel symbols on the boundaries introduce additional, localized contributions: 
\begin{subequations}\label{DeltaR}
\begin{align}
\Delta R^5_{\phantom{5}5\mu\nu} & = - \Delta R^5_{\phantom{5}\mu5\nu} \,=\, 2 k \, \warp{-2} \, \eta_{\mu\nu} \bigl[ \delta(y) - \delta (y - \pi R) \bigr] ; \\ 
\Delta R^\mu_{\phantom{\mu}5\nu5} & = - \Delta R^\mu_{\phantom{\mu}\nu55} \,=\, 2 k \, \delta^\mu_\nu \bigl[ \delta(y) - \delta (y - \pi R) \bigr] \,.
\end{align}
\end{subequations}
While these contributions are irrelevant for particles of lower spin (for example, all factors involving Christoffel symbols vanish in the Lagrangian for vector particles by antisymmetry), they have consequences for spin-two fields.

Finally, we note a convention we will use throughout: Roman indices $M,N,\ldots$ span the full five-dimensional space, and are raised and lowered with the full metric $g$, while Greek indices $\mu,\nu,\ldots$ span the four large dimensions only and are raised and lowered with the \emph{flat} metric $\eta_{\mu\nu}$.

\subsection{Kaluza-Klein Decomposition of Massive Spin-2 Field}

To model Regge excitations of the SM fields propagating in the RS bulk, we generalize the field theory of Section~\ref{sec:4D} in a straightforward way: First, we promote both the Regge and SM fields to five-dimensional fields, and introduce the appropriate metric factors, Christoffel symbols, etc. into the Lagrangian to restore general covariance. For SM fermions, bulk masses are introduced, and chiral 4D zero modes are obtained by imposing appropriate boundary conditions~\cite{GN,GP}. Then, we perform the Kaluza-Klein (KK) decomposition of the theory, and derive the interactions between the 4D fields. For spin-1 and spin-1/2 fields, the KK decomposition is straightforward. The KK decomposition for 
a massive spin-2 field in RS space is somewhat complicated, and to the best of our knowledge this problem has not yet been addressed in the literature. In this subsection, we will outline the required steps. 

A free (non-interacting) massive spin-2 field in curved 5D space is described by the covariant generalisation of~\eqref{Bin4D}:
\begin{equation}
\lagrange{L} = \frac{1}{4} \fst^{LMN} \fst_{LMN} - \frac{1}{2} \rselwr{\fst}{LM}{M} \lwrrse{\fst}{LN}{N} + \frac{1}{2} m^2 \left\{ \left( \lwrrse{\sptwo}{M}{M} \right)^2 - \sptwo^{MN} \sptwo_{MN} \right\} ,
\label{5Dflat}\end{equation}
where $\fst_{LMN} \equiv \nabla_L \sptwo_{MN} - \nabla_M \sptwo_{LN}$ is the field strength tensor. Under the 4D Lorentz group, the field decomposes into tensor, vector and scalar components, $B_{\mu\nu}$, $B_{\mu 5}$, and $B_{55}$, respectively. However, the Lagrangian~\leqn{5Dflat} contains terms which mix these components. To obtain a consistent KK decomposition, these mixed terms need to be cancelled. To do this, first note that in flat space, the kinetic part of Eq.~\eqref{5Dflat} is invariant (up to a total derivative) under the gauge transformation
\beq
\delta \sptwo_{MN} = \partial_M \beta_N + \partial_N \beta_M \,.
\eeq{GTflat}
The mass terms can be thought of as spontaneous breaking of this gauge invariance. One can formally restore the gauge invariance to this Lagrangian by introducing pion fields:\footnote{This idea is based on the non-linear sigma model for gravity constructed in Ref.~\cite{nlsmgrav}.}
\begin{multline}
\lagrange{L} = \frac{1}{4} \fst^{LMN} \fst_{LMN} - \frac{1}{2} \rselwr{\fst}{LM}{M} \lwrrse{\fst}{LN}{N} + \frac{1}{2} \left\{ \left( m \, \lwrrse{\sptwo}{M}{M} - 2 \partial_M \pi^M \right)^2 \right. \\
- \left( m \, \sptwo^{MN} - \partial^M \pi^N - \partial^N \pi^M \right) \bigl( m \, \sptwo_{MN} - \partial_M \pi_N - \partial_N \pi_M \bigr) \Bigr\} .
\label{withpions_flat}\end{multline}
If the pion fields transform as
\beq
\delta \pi_M = m\,\beta_M
\eeq{piontransform}
the Lagrangian~\leqn{withpions_flat} is gauge invariant. Setting the pion fields to zero corresponds to the unitary gauge, where the Lagrangian has the familiar form, Eq.~\leqn{5Dflat}. Following the standard prescription of the $R_\xi$ gauges, one can choose a gauge in which the mixings between fields of different 4D spins disappear. This gauge is a natural basis for KK decomposition.  

To apply this procedure in RS space, we must first determine the correct form of the gauge transformations, since the kinetic terms in~\leqn{5Dflat} are {\it not} invariant under~\leqn{GTflat} in the presence of curvature. We take the gauge transformations in warped space to be
\begin{align}
\delta \sptwo_{MN} & = \nabla_M \beta_N + \nabla_N \beta_M ; \notag \\
\delta \pi_M & = m \, \beta_M \,.
\label{GTcurve}\end{align}
Invariance under these transformations requires additional terms in the Lagrangian, which are proportional to curvature and disappear in flat-space limit. These terms are of two kinds: corrections to the bulk Lagrangian, and boundary-localized corrections. (The appearance of the boundary-localized terms is due to the boundary contributions to the curvature tensor, Eq.~\leqn{DeltaR}.) The bulk term is 
\beq
\Delta {\cal L}_{\rm bulk} = \frac{3k^2}{2} \left\{ \left( \lwrrse{\sptwo}{M}{M} \right)^2 - \sptwo^{MN} \sptwo_{MN} \right\}
 \,.
\eeq{DeltaLbulk}
This term agrees with that found in Ref.~\cite{Buchbinder:2000fy}.  The form of the boundary terms depends on the transformations of the $B_{MN}$ components under the action of the orbifold symmetry. We choose the 4D tensor field to be {\it even} under $y\rightarrow -y$; consistency then implies 
\beq
B_{\mu\nu}(y) = + B_{\mu\nu}(-y),~~~B_{\mu5}(y) = - B_{\mu5}(-y),
~~~B_{55}(y) = + B_{55}(-y)\,.
\eeq{bc_B}
These conditions in turn imply the following boundary conditions for the gauge functions $\beta_M$:
\beq
\partial_5 \left( \beta_\mu(x,y) e^{-2k|y|}\right)|_{y=0,\pi R} = 0,~~~
\beta_5(x,y)|_{y=0,\pi R} = 0\,.
\eeq{bc_beta}
To restore gauge invariance on the boundary, for gauge transformations satisfying~\leqn{bc_beta}, we add
\beq
\Delta {\cal L}_{\rm brane} = -k \left[ \delta(y) - \delta(y-\pi R)\right]  \left( \left( \lwrrse{\sptwo}{\mu}{\mu} \right)^2 - \sptwo^{\mu\nu} \sptwo_{\mu\nu}  
\right)\,.
\eeq{DeltaLbrane}
This term was not found in Ref.~\cite{Buchbinder:2000fy}, which did not consider possible brane-localized terms.  Summarizing, the gauge-invariant warped-space generalization of Eq.~\leqn{withpions_flat} is
\begin{multline}
\lagrange{L} = \frac{1}{4} \fst^{LMN} \fst_{LMN} - \frac{1}{2} \rselwr{\fst}{LM}{M} \lwrrse{\fst}{LN}{N} +  \frac{1}{2} \Bigl\{ \left( m \lwrrse{\sptwo}{M}{M} - 2 \nabla_M \pi^M \right)^2 \\
- \left( m \sptwo^{MN} - \nabla^M \pi^N - \nabla^N \pi^M \right) \bigl( m \sptwo_{MN} - \nabla_M \pi_N - \nabla_N \pi_M \bigr) \Bigr\} \\ + \Delta {\cal L}_{\rm bulk} + \Delta {\cal L}_{\rm brane}.
\label{5Dcorrect}\end{multline}
Finally, we comment on the connection between Eq.~\eqref{5Dcorrect} and general relativity.  It is well known that in curved spacetimes, the graviton Lagrangian is invariant under a gauge transformation of the form~\eqref{GTcurve} that arises from coordinate invariance.  Quadratic expansion of the Einstein-Hilbert action in curved space includes terms of the form $R \, h \, h$ (where $R$ is a curvature tensor) in the bulk, and $T\,h\,h$ (where $T$ is the brane tension, related to $R$ via Einstein's equations) on the branes. For the RS space, those terms are in fact identical to the terms added by hand in our approach, Eqs.~\eqref{DeltaLbulk} and~\eqref{DeltaLbrane}.

Most of the terms mixing fields of different 4D spin in this Lagrangian can be eliminated with judiciously chosen gauge-fixing terms. However, there are terms mixing spin-2 and spin-0 fields in~\eqref{5Dcorrect} that cannot be removed within the $R_\xi$-gauge approach; they have the form
\begin{multline}
\warp{-2} \bigl( -3 k \sgn (y) \, ( \partial_y \lwrrse{\sptwo}{\mu}{\mu} ) \sptwo_{yy} - (6 k^2 + m^2) \lwrrse{\sptwo}{\mu}{\mu} 
\sptwo_{yy} \\ + 2 m \lwrrse{\sptwo}{\mu}{\mu} \left( \partial_y \pi_y - 3k \sgn (y) \, \pi_y \right) \bigr) .
\label{spinorscalarmix}\end{multline}
This mixing can be eliminated by shifting the spin-2 field,
\begin{equation}
\sptwo_{\mu\nu} \to \sptwo_{\mu\nu} - \frac{1}{3} g_{\mu\nu} \phi \,.
\label{rotatespin2}\end{equation}
The shift $\phi$ is a function of the fields $\sptwo_{yy}$ and $\pi_y$, defined as a solution to the differential equation 
\begin{multline}
\left( - \partial_y^2 + 4 k^2 + m^2 \right) \, \warp{-2} \, \phi - 4k\,\bigl[ \delta(y) - \delta(y-\pi R) \bigr]  \warp{-2} \phi \\
= \warp{-2} \, \bigl( 3 k \sgn (y) \, \partial_y \sptwo_{yy} - (12 k^2 + m^2) \sptwo_{yy} + 2 m \, \partial_y \pi_y \\
- 6 m k \sgn (y) \, \pi_y \bigr) 
+ 6 k \, \bigl[ \delta(y) - \delta(y-\pi R) \bigr] \warp{-2} \sptwo_{yy} .
\label{defphi}\end{multline}
The operator acting on $\warp{-2} \, \phi$ arises from the terms that are quadratic in $\sptwo_{\mu\nu}$.  Since it is a self-adjoint operator with strictly positive eigenvalues, it can be inverted and thus $\phi$ exists. The shift must be done \emph{before} gauge fixing as it introduces terms mixing $\sptwo_{\mu\nu}$ and $\phi$ which must be cancelled by the gauge-fixing terms. 

To cancel the remaining mixing between the 4D tensor mode and 4D vectors and scalars, we introduce the gauge-fixing term
\begin{equation}
\lagrange{L}_\text{gf1} = \frac{1}{\xi} \warp{2} G^\mu G_\mu\,,
\label{gf1short}\end{equation}
where
\begin{multline}
G_\mu \equiv \warp{2} \left( \partial_\mu \lwrrse{\sptwo}{\nu}{\nu} - \partial^\nu \sptwo_{\mu\nu} \right) - \frac{1}{2} \xi \bigl( -2 \partial_y \sptwo_{\mu y} + 4 k \sgn (y) \, \sptwo_{\mu y} \\
+ 2 m \, \pi_\mu + \partial_\mu \sptwo_{yy} +  \frac{2}{3} \partial_\mu \phi \bigr)\,.
\label{gf1long}\end{multline}
This results in the action of the form
\begin{equation}
S = S_\text{spin-2} \oplus S_\text{spin-1, spin-0}  \,,
\label{decomposeaction}\end{equation}
where
\begin{multline}
S_\text{spin-2} = \int \ud^5 x \biggl[ \warp{2} \left( \frac{1}{4} \fst^{\lambda\mu\nu}\fst_{\lambda\mu\nu} - \frac{1}{2} \left( 1 - \frac{2}{\xi} \right) \rselwr{\fst}{\lambda\mu}{\mu} \lwrrse{\fst}{\lambda\nu}{\nu} \right) \\
+ \frac{1}{2} \lwrrse{\sptwo}{\mu}{\mu} \left( - \partial_y^2 + 4k^2 + m^2 \right) \lwrrse{\sptwo}{\nu}{\nu} - \frac{1}{2} \sptwo^{\mu\nu} \left( - \partial_y^2 + 4k^2 + m^2 \right) \sptwo_{\mu\nu} \\
+ \, 2k \left(\delta(y)-\delta(y-\pi R)\right) \left(\sptwo^{\mu\nu}\sptwo_{\mu\nu}\,-\,(\sptwo^\mu_\mu)^2 \right) \biggr]\,.
\label{5ds2action}\end{multline}
Additional gauge-fixing terms must be introduced to separate the vector and scalar fields in the action. Since the procedure is rather complicated, and since phenomenologically the tensor field provides the most interesting and unambiguous signature for stringy physics, we will not pursue a complete description of the vector and scalar sectors in this paper.

\begin{figure}
\centering
\includegraphics[width=0.6\textwidth]{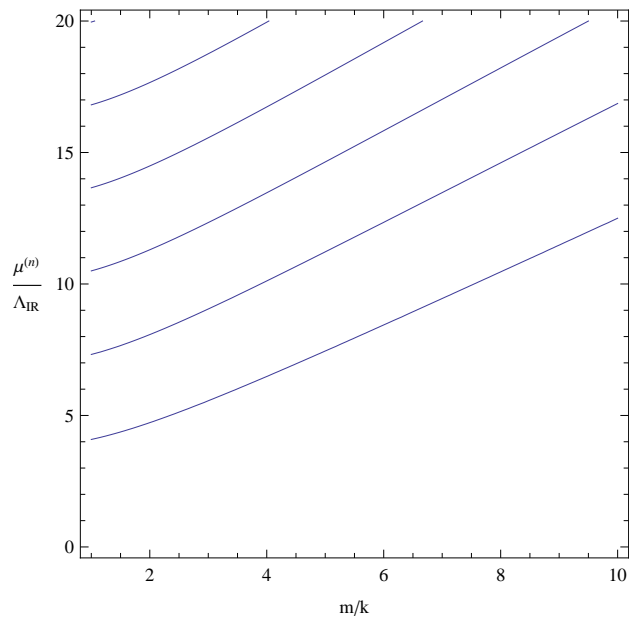}
\caption{The spectrum of 4D tensor particles.  We have assumed that the RS curvature $k = 10^{15}~$TeV; the results are essentially independent of this choice.}
\label{spin2massone}
\end{figure}

\begin{figure}
\centering
\includegraphics[width=0.8\textwidth]{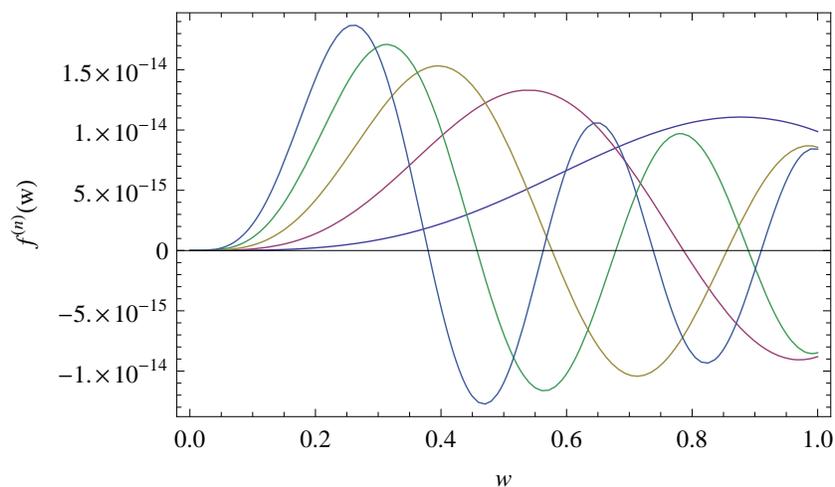}
\caption{The KK functions for the five lightest KK modes.  We have assumed that the RS curvature $k = 10^{15}~$TeV, $\LIR = 1$~TeV and $m = 3 k$.  The qualitative form of the wavefunctions are all robust to varying these parameters.}
\label{spin2wf}
\end{figure}

Once the tensor field is isolated in the action, KK decomposition is straightforward. We make the standard KK ansatz
\begin{equation}
\sptwo_{\mu\nu} (x,y) = \frac{1}{\sqrt{\pi R}} \, \sum_{n=1}^\infty \kkn{\sptwo_{\mu\nu}} (x) \kkn{f} (y) .
\label{tensorkk}\end{equation}
The defining equation for the $\{ f ^{(n)} \}$ are found easily from either the equations of motion or the action: 
\begin{equation}
- \kkn{f}{}'' + \left( 4k^2 + m^2 \right) \kkn{f} - 4k\left( \delta(y) - 
\delta(y-\pi R) \right) \kkn{f}= {\kkn{\mu}}^2 \warp{2} \kkn{f} .
\label{tensordefine}\end{equation}
This equation is self-adjoint, so we can take the KK functions to be orthonormal.  In this case, the associated  inner product is
\begin{equation}
\frac{1}{\pi R} \int_0^{\pi R} \ud y \, \warp{2} \kkn{f} \kkm{f} = \delta^{nm} .
\label{tensornorm}\end{equation}
After integrating over the extra dimension, the action~\eqref{5ds2action} becomes
\begin{multline}
S_\text{spin-2} = \int \ud^4 x \sum_{n=1}^\infty \biggl[ \frac{1}{4} \kkn{\fst}_{\lambda\mu\nu} \kkm{\fst}{}^{\lambda\mu\nu} - \frac{1}{2} \left( 1 - \frac{2}{\xi} \right) \rselwr{\fst}{(n)\lambda\mu}{\mu} \fst_{\lambda\nu}^{(n)\nu} \\
+ \frac{1}{2} {\mu^{(n)}}^2 \left\{ \left( \sptwo^{(n)\mu}_{\ \mu} \right)^2 - \sptwo^{(n)\mu\nu} \sptwo^{(n)}_{\mu\nu} \right\} \biggr] \,,
\end{multline}
which is just a tower of free 4D spin-2 fields with masses $\mu^{(n)}$ (in unitary gauge, if $\xi\rightarrow\infty$.)
The general solution to~\eqref{tensordefine} is a Bessel function:
\begin{equation}
\kkn{f} (y) = \frac{1}{N} \left\{ J_\nu \left( \frac{\kkn{\mu}}{\LIR} w \right) + c \, J_{-\nu} \left( \frac{\kkn{\mu}}{\LIR} w \right) \right\}\,,
\label{gensol}\end{equation}
where
\beq
\LIR = k e^{-\pi k R},~~~w = e^{k(|y|-\pi R)} \in [e^{-\pi k R}, 1]\,.
\eeq{defs_for_gensol}
The order of the Bessel function is $\nu \equiv \sqrt{4+\mathfrak{m}^2}$, where $\mathfrak{m}=m/k$ is the string scale in units of the RS curvature. Formally, consistent treatment of the RS geometry as a smooth background for propagating strings requires $\mathfrak{m} \gg 1$; in our phenomenological study, we will consider $\mathfrak{m}\sim$ a few. $N$ is the normalisation and $c$ is a constant of integration; each implicitly depends upon the level $n$.  Both $c$ and the mass are set by the boundary conditions.  Since $\sptwo_{\mu\nu}$ is even under the orbifold symmetry (see Eq.~\eqref{bc_B}) it (and hence the $\{\kkn{f}\}$) would normally satisfy Neumann boundary conditions.  However, the presence of localized terms in~\eqref{tensordefine} changes this, making the derivative of the KK functions discontinuous.  The correct boundary conditions are
\begin{subequations}
\begin{align}
\kkn{f}{}' (0+) - \kkn{f}{}' (0-) & = - 4k \kkn{f} (0) ;\\
\kkn{f}{}' (-\pi R+) - \kkn{f}{}' (\pi R-) & = 4k \kkn{f} (\pi R).
\end{align}
\end{subequations}
We plot the spectrum that is implied by these boundary conditions in Fig.~\ref{spin2massone}, and the numerical values of the lightest tensor Reggeon mass for a few choices of $m/k$ are listed in Table~\ref{tab:Numbers}. The wavefunctions of the first five modes, for a specific choice of $m/k=3$, are plotted in Fig.~\ref{spin2wf}. As expected, the mass of the spin-2 Reggeon (and its first few KK excitations) is of the order
\beq
\mu \sim~{\rm (a~few)}\times me^{-\pi k R}\,\sim~{\rm a~few~TeV}\,,
\eeq{mass}
and the wavefunctions are strongly localized in the vicinity of the IR brane, $y=\pi R$. It is also easy to roughly estimate the two constants appearing in the wavefunctions: 
\beq
N \sim \frac{1}{\sqrt{\pi kR}} e^{+\pi kR}\,,~~~c \sim e^{-2\nu \pi kR}\,.
\eeq{estimates}
These estimates are useful in discussing the Reggeon phenomenology. 

\section{SM Couplings to the Tensor Reggeon}

To model the interactions of SM quarks and gluons with the Regge gluon, we will use the minimal generally covariant extension of the interaction Lagrangian of the 4D, flat-space CPP model, discussed in Section 2.

\subsection{Gluon-Reggeon Coupling}

The gluon-Reggeon coupling is a simple generalization of Eq.~\leqn{Lqcd4gluon}:
\beq
{\cal S}_{ggg^*} \,=\, \int d^5x \sqrt{-G} \,\frac{g_5}{\sqrt{2}M_S^*}\,C^{abc}\,\left(F^{aAC} F^{bB}_C\,-\frac{1}{4}\,F^{aCD}F^b_{CD}G^{AB}\right)\, B^c_{AB}\,. 
\eeq{Lggg5}
The 5D coupling $g_5$ is related to the 4D QCD coupling $g_s$ by
\beq
g_5 = \sqrt{\pi R} \, g_s\,.
\eeq{couplings}
The 5D gauge coupling has mass dimension of $-1/2$, so that the power-counting of the operator~\leqn{Lggg5}
is correct for canonically normalized fields (namely, the 5D gauge field and the Reggeon both have mass dimension 3/2).    
The interaction Lagrangian is invariant under the usual QCD gauge transformations (see discussion in section~\ref{sec:stringyQCD}), but not under the transformations~\leqn{GTcurve}. It is easy to formally restore this symmetry by replacing $B_{MN}\rightarrow B_{MN}-\nabla_M \pi_N - \nabla_N \pi_M$; however, the terms involving pions do not contribute to the couplings of the 4D tensor mode, which is the only object of interest for us.   

The Kaluza-Klein decomposition of the SM gauge field is straightforward~\cite{DHR}. Gauge freedom can be used to choose $A_5=0$, and the 4D vector zero-mode has a constant profile in the bulk:
\beq
A_\mu(x,y) = \frac{1}{\sqrt{\pi R}} A_\mu^{(0)}(x) + \ldots\,.
\eeq{A0mode}
This yields the following 4D Lagrangian for the interactions of the SM gluons with the tensor Reggeons:
\beq
{\cal L}_{ggg^*} \,=\, \sum_n \frac{g^{(n)}}{\sqrt{2}\tilde{M}_S} {\cal C}^{abc} \left( F^{a\alpha\gamma}F^{b\,\,\beta}_\gamma\,-\,\frac{1}{4}F^{a\gamma\delta}F^b_{\gamma\delta}\right) \, B^c_{\alpha\beta}\,,
\eeq{4Dglue}
where we defined the warped-down string scale
\beq
\tilde{M}_S \,=\, e^{-\pi k R} M_S^*\,\sim\,{\rm a~few~TeV}, 
\eeq{Mtilde}
and the dimensionless coupling
\beq
g^{(n)} = \frac{g_s\,e^{-\pi k R}}{\pi R} \int_0^{\pi R} dy \, e^{2ky} \, f^{(n)} (y)\,.  
\eeq{gluevertex}
Since the Reggeon wavefunction is localized near $y=\pi R$, and is of order $1/N\sim \sqrt{\pi kR} e^{-\pi kR}$ in that region, we can estimate
\beq
g^{(n)} \sim \frac{g_s}{\sqrt{\pi kR}} \, .
\eeq{gnestimate} 
The operator in the 4D action, Eq.~\leqn{4Dglue}, is suppressed by a scale of order $\tilde{M}_S$, as expected; however, note the additional volume suppression. Sample numerical values for the coupling of the lightest Reggeon $g^{(0)}$ are shown in Table~\ref{tab:Numbers}. 
The coupling is approximately independent of the Reggeon mass, with a value of roughly $0.1g_s$.

\subsection{Quark-Reggeon Coupling}

Embedding of the SM fermions as zero modes of 5D fermions in the RS background is well known~\cite{GN,GP}. For each SM chiral quark, we introduce a 5D field $Q_i$, where the index $i$ includes both chirality and flavor. Generalizing Eq.~\leqn{Lqcd4}, the Regge gluon couples to this field via
\begin{multline}
{\cal S}_{q\bar{q}g^*} \,=\, \int d^5x \sqrt{-G} \,\frac{i g_5}{\sqrt{2}M_S^*}\,G^{LM}\,E^N_a\,
\left( (\overline{{\cal D}_M Q_i})\Gamma^a \tilde{\sptwo}_{LN} Q_i - 
\overline{Q}_i \Gamma^a  \tilde{\sptwo}_{LN} {\cal D}_M Q_i \right) \,,
\label{Lqqg5}\end{multline}
where $\Gamma^n=(\gamma^\nu, i\gamma^5)$, and $
E^N_n(y) ={\rm ~diag~}(e^{k|y|},e^{k|y|},e^{k|y|},e^{k|y|},1)$
is the inverse vierbein. The covariant derivative has the form (up to terms containing gauge fields)
\beq
{\cal D}_M Q  = \partial_M Q + \frac{1}{2}\,\omega^{ab}_M \sigma_{ab}\,,
\eeq{covder} 
where $\omega^{ab}$ is the spin connection, and $\sigma_{ab} = 
\frac{1}{4}[\Gamma_a, \Gamma_b]$. (Note that the indices $a,b,\ldots$ refer to the transformations under local Lorentz group, and as such are raised and lowered with Minkowski metric.) In RS space, the only non-vanishing components of the spin connection are
\beq
\omega^{\alpha 5}_\mu = -\omega^{5\alpha}_\mu = -k~{\rm sgn}(y) 
\,e^{-k|y|}\,\delta^\alpha_\mu\,.
\eeq{spincon}
It is easy to show that the terms involving spin connection in the action~\leqn{Lqqg5} are proportional to the trace of the tensor Reggeon, $B^\mu_\mu$, and thus vanish for an on-shell Reggeon. As a result, the covariant derivatives in Eq.~\leqn{Lqqg5} can be replaced with ordinary derivatives when considering an on-shell Reggeon, as will always be the case in this paper.

The zero-mode quarks $q_i(x)$ are given by~\cite{GN,GP}
\beq
Q_i(x,y) = N_i e^{(2-c_i)k|y|} q_i(x) + \ldots
\eeq{Q0modes}
where the normalization constant is
\beq
N_i = k^{1/2} \sqrt{\frac{1-2c_i}{e^{\pi k R (1-2c_i)} - 1}}\,.
\eeq{norms}
The 4D fields $q_i$ are canonically normalized.
The $c$ parameters are related to the 5D fermion masses $M_5$: 
in the notation of Ref.~\cite{GN}, $c=M_5/k$ for the right-handed fields 
and $c=-M_5/k$ for the left-handed fields. (The ``handedness'' of the 5D fields refers to the 4D chirality of their zero modes.) We work in the basis where the bulk masses are diagonal.\footnote{The SM fermion masses and mixings may be due to the interactions of the bulk fermions with a brane-localized Higgs boson~\cite{ADMS}, or, in Higgsless models, to modified boundary conditions~\cite{CGHST}. We will not consider these effects in this paper, since they do not have a major effect on the Reggeon collider phenomenology.} In four dimensions, we obtain
\beq
{\cal L}_{q\bar{q}g^*} \,=\,\sum_n \frac{i \tilde{g}_i^{(n)}}{\sqrt{2}\tilde{M}_S}
\left(\partial^\mu \bar{q} \gamma^\nu \tilde{B}_{\mu\nu} q \,-\,
\bar{q} \gamma^\nu \tilde{B}_{\mu\nu} \partial^\mu q \right)\,,
\eeq{4Dquarks}
where
\beq
\tilde{g}_i^{(n)} \,=\, g_s e^{-\pi kR} N_i^2 \int_0^{\pi R} \,dy\,f^{(n)}(y)\,e^{(3-2c_i)
k|y|}\,.
\eeq{gtilde_def}
Clearly, the strength of the coupling depends crucially on the value of $c_i$. As a rough estimate, we obtain
\begin{subequations}
\begin{align}
\tilde{g}_i & \sim g_s\,e^{(1-2c_i)k\pi R},~~~c_i > \frac{1}{2}\,; \\
\tilde{g}_i  & \sim \frac{g_s}{\sqrt{\pi k R}} \approx \frac{g_s}{6},~~~c_i = \frac{1}{2}\,; \\
\tilde{g}_i & \sim g_s,~~~c_i < \frac{1}{2}\,.
\end{align}
\end{subequations}
Thus, the couplings to ``elementary" fermions ($c>1/2$) are exponentially suppressed, couplings to ``mixed" fermions ($c=1/2$) are volume-suppressed, and couplings to ``composite" fermions ($c<1/2$) are unsuppressed. This behavior is consistent with naive expectations from the dual CFT picture, where the Reggeon is a composite.  
Numerical values for the couplings of the lightest Reggeon to fermions with three sample values of $c$ are shown in Table~\ref{tab:Numbers}. 

\begin{table}[t!]
\begin{center}
\begin{tabular}{|r||r|r|r|r|r|} \hline
$\frac{m}{k}$ & $\frac{\mu^{(0)}}{\Lambda_{\rm IR}}$ & 
$\frac{g^{(0)}}{g_s}$ 
& $\frac{\tilde{g}^{(0)}}{g_s}(c=0.65)$ & $\frac{\tilde{g}^{(0)}}{g_s}(c=0.5)$ &  $\frac{\tilde{g}^{(0)}}{g_s}(c=0)$ 
\rule{0ex}{2.2ex} \\
\hline
2.0 & 4.72 & 0.110 & $3.9\times 10^{-5}$ & 0.110 & 2.9\\ 
3.0 & 5.56 & 0.107 & $3.8\times 10^{-5}$ & 0.107 & 2.9 \\
4.0 & 6.48 & 0.104 & $3.7\times 10^{-5}$ & 0.104 & 2.9 \\
5.0 & 7.45 & 0.100 & $3.5\times 10^{-5}$ & 0.100 & 2.8 \\
\hline 
\end{tabular} \\[1ex]
\caption{Mass of the lightest tensor Reggeon and its couplings to  gluons and quarks (with three different values of $c$), as a function of the bulk Reggeon mass $m$. We have assumed $k/\LIR = 10^{15}$; masses and couplings only depend on $k$ and $R$ through a logarithm of this ratio.}
\label{tab:Numbers}
\end{center}
\end{table}

The values of $c_i$ for various quark flavors are somewhat model-dependent. We will study the 
Reggeon phenomenology in two scenarios. The first one is the model with a light Higgs on the IR brane~\cite{ADMS}. If the brane-localized Yukawa couplings are anarchic, the SM pattern of masses and mixings leads to the following estimates for these coefficients~\cite{APS}:
\begin{subequations}
\begin{align}
c_{Q^1} &\approx 0.63,~~~c_{u^1}\approx 0.675,~~~c_{d^1}\approx
0.675\,;\\
c_{Q^2} &\approx 0.575,~~~c_{u^2}\approx 0.5,~~~c_{d^2}\approx
0.64\,;\\ 
c_{Q^3} &\approx 0.39,~~~c_{u^3}\approx -0.19,~~~c_{d^3}\approx
0.62\,.
\label{cs}
\end{align}
\end{subequations}
In this scenario, the first two generations of quarks are mostly elementary, and their couplings to the tensor Reggeon are exponentially suppressed (numerically, the suppression factor is of order $10^{-5}-10^{-6}$). The couplings to the third generation doublet and the right-handed top quark are unsuppressed.

Our second scenario is the``Higgsless" model~\cite{HL,CGHST}. In this model, consistency with precision electroweak constraints requires~\cite{HLdeloc}
\beq
c_i \approx \frac{1}{2}
\eeq{HLc} 
for all flavors, with the exception of the third-generation doublet and the right-handed top, which have approximately the same $c$ values as in Eq.~\leqn{cs}.  In this scenario, the tensor Reggeon couples to light quarks, with a coupling suppressed only by the volume factor. 

\section{Phenomenological Implications}

The most important parameter that determines the sensitivity of the LHC experiments to a new particle is its mass. The tensor Reggeon mass in our model depends on two parameters, $\LIR$ and $m/k$; for fixed $m/k$, the Reggeon mass is, to an excellent approximation, a linear function of $\LIR$. The scale $\LIR$ is subject to a number of significant constraints from existing experiments. Bounds from precision electroweak measurements and flavor physics have been considered by many authors, both in models with the Higgs and in the Higgsless approach. 
Among these, precision electroweak bounds, in particular the bound from the $S$ parameter, are considered to be more robust, since no  
known symmetry can be used to avoid it. In the model with the Higgs, the bound on the first KK excitation mass is of order 3 TeV~\cite{ADMS}, translating into roughly $\LIR\gsim 1$ TeV. In the Higgsless model, the KK excitations of the $W$ bosons must lie below 1 TeV for unitarity, corresponding to $\LIR\lsim 0.5$ TeV. This is only consistent with precision electroweak constraints if all SM fermions, with the exception of the right-handed top quark, have approximately flat profiles in the extra dimension, $c_i\approx 1/2$~\cite{HLdeloc}.
Since lower KK masses generally require more finely-tuned fermion profiles, we will adapt the value $\LIR= 0.5$ TeV for this model. As we already remarked in the Introduction, the description of physics in the RS model as strings propagating on a smooth geometric background formally requires $m/k\gg 1$; however, as in many examples in various areas of physics, $m/k\sim$ a few may in fact be sufficient, depending on the behavior of the leading corrections to the geometric limit, as well as on desired accuracy. Precise determination of the domain of validity of geometric description is beyond the scope of this paper. The lower the allowed value of $m/k$, the lighter the tensor Reggeon can be; for example, assuming that $m/k\geq3$ is acceptable, we find that the lowest tensor Reggeon mass is about $5\LIR$ (see Fig.~\ref{spin2massone}), corresponding to 2.5 TeV in the Higgsless model and above 5 TeV in the model with the Higgs.

\begin{figure}
\centering
\includegraphics[width=0.8\textwidth]{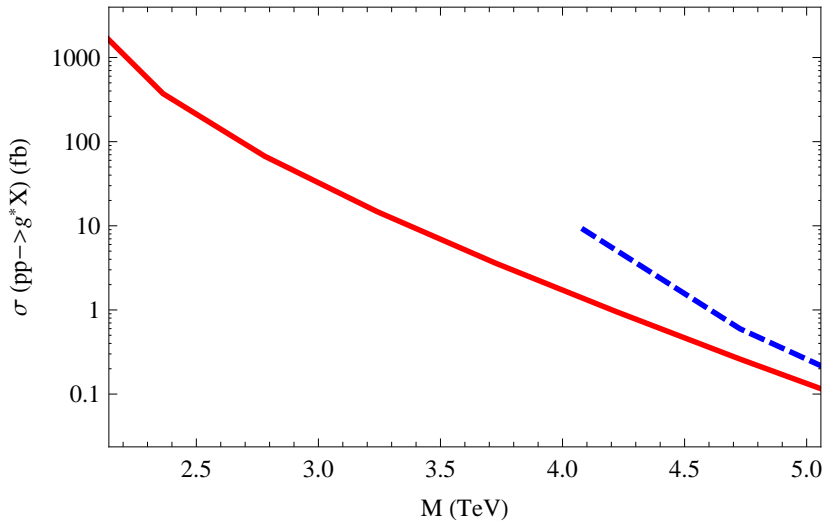}
\caption{Production cross section of the lowest-lying tensor Regge gluon at the LHC, $\sqrt{s}=14$ TeV. Red/solid line: Higgsless model. Blue/dashed line: Model with a Higgs. (See text for detailed definition of the two models.)}
\label{fig:xsec}
\end{figure}

The second crucial quantity for experimental searches for the tensor Reggeon is its production cross section. For the lightest Reggeon, parton-level cross sections are given by 
\begin{subequations}
\begin{align}
\hat{\sigma}(q_i \bar{q}_i \rightarrow g^*) & = \, \frac{2\pi^2\alpha_s}{9}\,\left( \frac{\tilde{g}^{(0)}_i M}{g_s \tilde{M}_S} \right)^2\,\delta(\hat{s}-M^2)\,; \\
\hat{\sigma}(g g \rightarrow g^*) & = \,\frac{5\pi^2\alpha_s}{6}\, \left( \frac{g^{(0)} M}{g_s \tilde{M}_S} \right)^2\,\delta(\hat{s}-M^2)\,,
\end{align}
\end{subequations}
where $M\equiv \mu^{0}$ is the Reggeon mass, and $q_i$ are Weyl (2-component) SM quarks. (Note that in the model with the Higgs, Regge gluon couplings to light quarks violate parity due to different 5D profiles of left-handed and right-handed SM quarks.)
The total production cross section at the LHC ($\sqrt{s}=14$ TeV), evaluated using the MSTW NLO parton distribution function set~\cite{MSTW}, is shown as a function of the Reggeon mass in Fig.~\ref{fig:xsec}. In this plot, we have assumed $m=M_S^*$ (as is in fact required for our Lagrangian to reproduce the Veneziano amplitudes in the flat-space limit). We further assumed $\LIR=1$ TeV in the model with the Higgs, $\LIR=500$ GeV in the Higgsless model, and $k/\LIR=10^{15}$ in both models. The range of Reggeon masses plotted in Fig.~\ref{fig:xsec} corresponds to $m/k\gsim 1$; we remind the reader that the results for small $m/k$ should be interpreted with caution since our framework may not be applicable.

We conclude that a significant sample (possibly thousands or even tens of thousands) of tensor Regge gluons could be produced at the LHC, for favorable model parameters.  The Reggeon production cross sections is similar to that of a KK gluon~\cite{KKgluonLHC} in the $2-3$ TeV range, but decreases faster with mass. Maximum production cross sections are of order a few pb in the Higgsless model, and 10 fb in the model with the Higgs. Note, however, that the lower cross section in the model with the Higgs is just due to the higher value of $\LIR$ assumed for that model. While this value is suggested as a lower bound by precision electroweak constraints, it could in principle be lowered at a price of fine-tuning, in which case lower Reggeon mass, and higher production cross section, would be possible. For the same value of the Reggeon mass, the model with the Higgs in fact predicts a somewhat higher production cross section than the Higgsless model, primarily due to a lower value of $\tilde{M}_S$.


\begin{figure}[t!]
\centering
\includegraphics[width=0.8\textwidth]{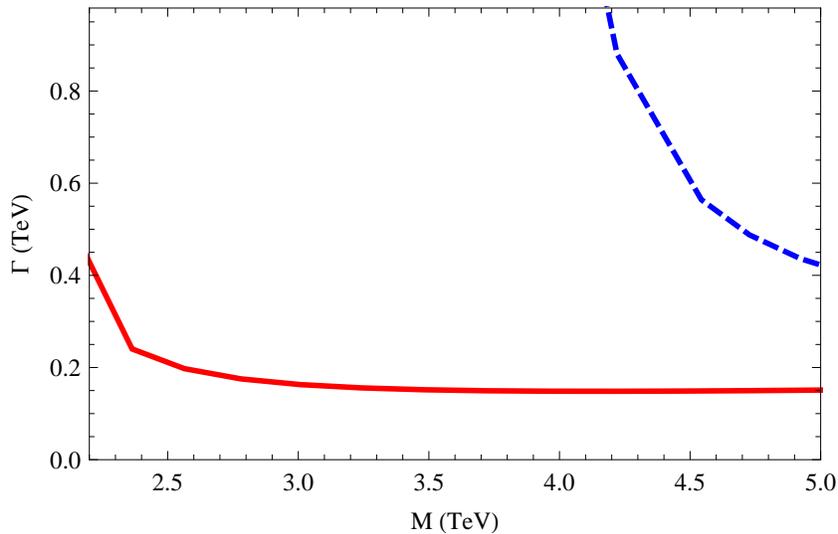}
\caption{Decay width of the lowest-lying Reggeon as a function of its mass. Red/solid line: Higgsless model. Blue/dashed line: Model with a Higgs. (See text for detailed definition of the two models.)}
\label{fig:width}
\end{figure}

Finally, the experimental signatures of the Reggeon production depend on its decay pattern. The partial widths are\footnote{In the flat space limit, our formulas agree with the corresponding results of Ref.~\cite{flatwidths}. We are grateful to Tomasz Taylor for pointing out a factor of 2 error in the gluon partial width in the original version of this paper.}
\begin{subequations}
\begin{align}
\Gamma(g^* \rightarrow q_i \bar{q}_i) & = \, \frac{\alpha_s\,M}{40}\,\left( \frac{\tilde{g}^{(0)}_i M}{g_s \tilde{M}_S} \right)^2\,\,; \\
\Gamma(g^*\rightarrow g g) & = \,\frac{\alpha_s\,M}{6}\, \left( \frac{g^{(0)} M}{g_s \tilde{M}_S} \right)^2\,,
\end{align}
\end{subequations}
where, as before, $q_i$ are two-component quarks. The total width of the Reggeon into SM channels is shown in Fig.~\ref{fig:width}. Among the SM channels, decays to top-antitop pairs dominate in both models under consideration: the branching ratio into tops (assuming that only SM decay channels are open) is about 95\% throughout the interesting mass range. Right-handed tops are preferred. Since the Reggeon mass is expected to be in the few TeV range, the tops would be highly boosted in the lab frame, so that the top decay products are strongly collimated into``top jets". Experimental and theoretical issues related to distinguishing such jets from light quark/gluon jets have been analyzed recently in a number of papers, in the context of KK gluon searches~\cite{topjets}. The proposed techniques would apply to Regge gluon searches as well. Since the Reggeon momentum can be fully reconstructed in events with hadronic top decays, such events could in principle be used to determine the angular distribution of the  
tops with respect to the beam axis, which would in turn allow one to determine the spin of the Reggeon and unambiguously distinguish it from a KK gluon.

In addition to the SM decays, the Reggeon may decay to other exotic states. For example, if the Reggeon mass is large enough, it can decay into pairs of Kaluza-Klein excitations of the SM quarks and gluons, which would in turn decay down to SM particles. We will not attempt to analyze such cascade decays in this paper. 

\section{Conclusions}

In this paper, we constructed a field-theoretic toy model to describe the lowest-lying Regge excitations of the SM gauge bosons, in a framework of the Randall-Sundrum model with all SM fields propagating in the bulk. We focused on the 4D tensor (spin-2) states, which would provide a clear signature of the underlying stringy physics if discovered. Our toy model allows us to predict the spectrum of these states (as a function of the underlying model parameters, including the fundamental string scale $M^*_S$), and their on-shell couplings to Standard Model fermions and gauge bosons. This is sufficient to make predictions for the processes that would dominate the Reggeon phenomenology at the LHC. 

If the curvature of the RS space is taken to zero, our model by construction reproduces the spectrum and couplings of the Reggeons in the toy model of CPP~\cite{CPP}. The CPP results were derived by  factorizing Veneziano amplitudes of string theory on $s$-channel poles. While the embedding of the SM into string theory in the CPP model was hardly fully realistic, the Veneziano amplitudes, and the on-shell Reggeon couplings derived from them, are very generic and do not depend on many of the details of realistic string compactifications. On the other hand, one should keep in mind that the Reggeon couplings in curved space may contain additional operators which vanish in the flat-space limit. A drawback of our approach is that it has no sensitivity to such operators, and while we do not expect their presence to result in qualitative changes to the picture obtained in our model, order-one numerical corrections seem possible. To properly handle this issue, one would need to properly quantize full string theory on the RS background, study four-particle scattering amplitudes, and factorize them to obtain Reggeon interaction vertices. 

Our model can be extended in a number of ways. First, 4D vector and scalar excitations of the SM vector bosons, as well as Regge excitations of the SM fermions, can be included. Second, higher Regge levels can in principle be considered, although it seems very unlikely that those could be within the LHC range. The contributions of the Reggeons to precision electroweak and flavor observables can be computed within our model, and may lead to additional constraints on models of this type. We leave all these issues for future work.

\vskip0.5cm

\noindent{\large \bf Acknowledgments} 

\vskip0.5cm

We are grateful to Lisa Randall for stimulating this investigation. We thank Csaba Csaki, Sohang Gandhi, Liam McAllister, and Matthew Reece for useful discussions. This research is supported by the U.S. National Science Foundation through grant PHY-0757868 and CAREER award PHY-0844667.


\begin{thebibliography}{99}

\bibitem{ADD}
  N.~Arkani-Hamed, S.~Dimopoulos and G.~R.~Dvali,
  Phys.\ Lett.\  B {\bf 429}, 263 (1998)
  [arXiv:hep-ph/9803315]; \\
  I.~Antoniadis, N.~Arkani-Hamed, S.~Dimopoulos and G.~R.~Dvali,
  Phys.\ Lett.\  B {\bf 436}, 257 (1998)
  [arXiv:hep-ph/9804398]; \\
   N.~Arkani-Hamed, S.~Dimopoulos and G.~R.~Dvali,
  Phys.\ Rev.\  D {\bf 59}, 086004 (1999)
  [arXiv:hep-ph/9807344].
  
\bibitem{RS}
  L.~Randall and R.~Sundrum,
  Phys.\ Rev.\ Lett.\  {\bf 83}, 3370 (1999)
  [arXiv:hep-ph/9905221].

\bibitem{CPP}
  S.~Cullen, M.~Perelstein and M.~E.~Peskin,
  Phys.\ Rev.\  D {\bf 62}, 055012 (2000)
  [arXiv:hep-ph/0001166].

\bibitem{Hunter}
  D.~Lust, S.~Stieberger and T.~R.~Taylor,
  Nucl.\ Phys.\  B {\bf 808}, 1 (2009)
  [arXiv:0807.3333 [hep-th]].

\bibitem{Regge_others}
  P.~Burikham, T.~Figy and T.~Han,
  Phys.\ Rev.\  D {\bf 71}, 016005 (2005)
  [Erratum-ibid.\  D {\bf 71}, 019905 (2005)]
  [arXiv:hep-ph/0411094];\\
  L.~A.~Anchordoqui, H.~Goldberg, S.~Nawata and T.~R.~Taylor,
  Phys.\ Rev.\ Lett.\  {\bf 100}, 171603 (2008)
  [arXiv:0712.0386 [hep-ph]];\\  
  L.~A.~Anchordoqui, H.~Goldberg, S.~Nawata and T.~R.~Taylor,
  Phys.\ Rev.\  D {\bf 78}, 016005 (2008)
  [arXiv:0804.2013 [hep-ph]];\\
  L.~A.~Anchordoqui, H.~Goldberg, D.~Lust, S.~Nawata, S.~Stieberger and T.~R.~Taylor,
  Phys.\ Rev.\ Lett.\  {\bf 101}, 241803 (2008)
  [arXiv:0808.0497 [hep-ph]].
  
  
  
\bibitem{Lust_review}
  D.~Lust,
  JHEP {\bf 0903}, 149 (2009)
  [arXiv:0904.4601 [hep-th]].

\bibitem{RSgut}
  L.~Randall and M.~D.~Schwartz,
  JHEP {\bf 0111}, 003 (2001)
  [arXiv:hep-th/0108114];\\
   L.~Randall and M.~D.~Schwartz,
  Phys.\ Rev.\ Lett.\  {\bf 88}, 081801 (2002)
  [arXiv:hep-th/0108115];\\
  W.~D.~Goldberger and I.~Z.~Rothstein,
  Phys.\ Rev.\ Lett.\  {\bf 89}, 131601 (2002)
  [arXiv:hep-th/0204160];\\
  W.~D.~Goldberger and I.~Z.~Rothstein,
  Phys.\ Rev.\  D {\bf 68}, 125011 (2003)
  [arXiv:hep-th/0208060];\\
  K.~Agashe, R.~Contino and R.~Sundrum,
  Phys.\ Rev.\ Lett.\  {\bf 95}, 171804 (2005)
  [arXiv:hep-ph/0502222].

\bibitem{ADMS}
  K.~Agashe, A.~Delgado, M.~J.~May and R.~Sundrum,
  JHEP {\bf 0308}, 050 (2003)
  [arXiv:hep-ph/0308036].

\bibitem{GN}
  Y.~Grossman and M.~Neubert,
  Phys.\ Lett.\  B {\bf 474}, 361 (2000)
  [arXiv:hep-ph/9912408].
  
\bibitem{APS}
  K.~Agashe, G.~Perez and A.~Soni,
  Phys.\ Rev.\  D {\bf 71}, 016002 (2005)
  [arXiv:hep-ph/0408134].
  
\bibitem{HL}
  C.~Csaki, C.~Grojean, L.~Pilo and J.~Terning,
  Phys.\ Rev.\ Lett.\  {\bf 92}, 101802 (2004)
  [arXiv:hep-ph/0308038].

\bibitem{Meade}
  P.~Meade and L.~Randall,
  JHEP {\bf 0805}, 003 (2008)
  [arXiv:0708.3017 [hep-ph]].

\bibitem{HMRR}
  B.~Hassanain, J.~March-Russell and J.~G.~Rosa,
  arXiv:0904.4108 [hep-ph].

\bibitem{fierzpauli}
  M.~Fierz and W.~Pauli,
  Proc.\ Roy.\ Soc.\ Lonf.\ A \textbf{173}, 211 (1939)

  
 \bibitem{GP} 
  T.~Gherghetta and A.~Pomarol,
  Nucl.\ Phys.\  B {\bf 586}, 141 (2000)
  [arXiv:hep-ph/0003129];\\
  S.~J.~Huber and Q.~Shafi,
  Phys.\ Lett.\  B {\bf 498}, 256 (2001)
  [arXiv:hep-ph/0010195].
    
\bibitem{nlsmgrav}
  N.~Arkani-Hamed, H.~Georgi and M.~D.~Schwartz,
  Ann.\ Phys.\ \textbf{305}, 96 (2003)
  [arXiv:hep-th/0210184].

\bibitem{Buchbinder:2000fy}
  I.~L.~Buchbinder, D.~M.~Gitman and V.~D.~Pershin,
  Phys.\ Lett.\  B {\bf 492}, 161 (2000)
  [arXiv:hep-th/0006144].

\bibitem{DHR}
  H.~Davoudiasl, J.~L.~Hewett and T.~G.~Rizzo,
  Phys.\ Lett.\  B {\bf 473}, 43 (2000)
  [arXiv:hep-ph/9911262].
  
  
\bibitem{CGHST}
  C.~Csaki, C.~Grojean, J.~Hubisz, Y.~Shirman and J.~Terning,
  Phys.\ Rev.\  D {\bf 70}, 015012 (2004)
  [arXiv:hep-ph/0310355].
  
\bibitem{HLdeloc}
  G.~Cacciapaglia, C.~Csaki, C.~Grojean and J.~Terning,
  Phys.\ Rev.\  D {\bf 71}, 035015 (2005)
  [arXiv:hep-ph/0409126].
  
\bibitem{MSTW}
  A.~D.~Martin, W.~J.~Stirling, R.~S.~Thorne and G.~Watt,
  arXiv:0901.0002 [hep-ph].
  
\bibitem{KKgluonLHC}
  K.~Agashe, A.~Belyaev, T.~Krupovnickas, G.~Perez and J.~Virzi,
  Phys.\ Rev.\  D {\bf 77}, 015003 (2008)
  [arXiv:hep-ph/0612015];\\
  B.~Lillie, L.~Randall and L.~T.~Wang,
  JHEP {\bf 0709}, 074 (2007)
  [arXiv:hep-ph/0701166].
  
\bibitem{flatwidths}
  L.~A.~Anchordoqui, H.~Goldberg and T.~R.~Taylor,
  Phys.\ Lett.\  B {\bf 668}, 373 (2008)
  [arXiv:0806.3420 [hep-ph]].
  
\bibitem{topjets}  
  J.~Thaler and L.~T.~Wang,
  JHEP {\bf 0807}, 092 (2008)
  [arXiv:0806.0023 [hep-ph]];\\
  D.~E.~Kaplan, K.~Rehermann, M.~D.~Schwartz and B.~Tweedie,
  Phys.\ Rev.\ Lett.\  {\bf 101}, 142001 (2008)
  [arXiv:0806.0848 [hep-ph]];\\
  L.~G.~Almeida, S.~J.~Lee, G.~Perez, G.~Sterman, I.~Sung and J.~Virzi,
  Phys.\ Rev.\  D {\bf 79}, 074017 (2009)
  [arXiv:0807.0234 [hep-ph]].
    
\end{thebibliography}
\end{document}